\documentclass[twocolumn,english,aps,prl]{revtex4-2}
\usepackage[T1]{fontenc}
\usepackage[latin9]{inputenc}
\usepackage{geometry}
\usepackage{color}
\usepackage{bm}
\usepackage{amsmath}
\usepackage{graphicx}
\usepackage{babel}
\begin{document}
\title{Matrix phase-space representations for gaussian boson sampling}
\author{Peter D. Drummond, Alexander S. Dellios, Margaret D. Reid}
\affiliation{Centre for Quantum Science and Technology Theory, Swinburne University
of Technology, Melbourne 3122, Australia}
\email{peterddrummond@protonmail.com}

\begin{abstract}
We introduce coherent matrix phase-space distributions. These use
conservation laws and symmetries to improve the accuracy and speed
of quantum phase-space representations. As an example, this is applied
to validation of low-loss Gaussian boson sampling (GBS) quantum computational
advantage experiments, where classical generation of the random photon-number
counts is exponentially hard. Large improvements in sampling errors
are demonstrated compared to previous methods. Matrix phase-space
representations also provide a large numerical speed-up, due to their
(at worst) quadratic scaling, compared to other methods for validating
total count probabilities of large-scale, low-loss GBS networks.
\end{abstract}
\maketitle
\textbf{Introduction}: Phase-space representations have a long history
in quantum theory \citep{Wigner_1932,Husimi1940}, and were applied
to quantum optics using the normally-ordered Glauber-Sudarshan P-representation
\citep{Glauber_1963_P-Rep,Sudarshan_1963_P-Rep}. Here, we develop
matrix phase-space representations which include symmetries in the
basis, thereby leveraging symmetry theory to make phase-space methods
more efficient. The resulting representation is complete, and has
greatly reduced sampling errors when symmetries exist. This uses superpositions
of coherent Schr\"{o}dinger cat states \citep{schrodinger1935gegenwartige},
allowing one to include conserved quantities. We treat a phase symmetry
in this Letter, but the method holds for more general cases. Our approach
unifies normally-ordered positive-P (+P) \citep{drummond1980generalised}
and gauge-P representations \citep{Deuar:2002} with stochastic wave-function
methods \citep{Carusotto:2001}.

To illustrate this, a simulation of moments and grouped probabilities
for low-loss Gaussian boson sampling (GBS) quantum computers (QC)
with photon-number resolving (PNR) detectors \citep{Aaronson2011,AaronsonArkhipov2013LV,Hamilton2017gaussian}
is demonstrated for up to 8000 modes. GBS is a non-universal optical
QC that sends squeezed photons into a large interferometer. Output
photons are measured, making a physical random number generator (RNG)
whose probabilities would require computing the \#P-hard Hafnian matrix
function \citep{Hamilton2017gaussian}. For large systems, exact simulation
of the counts is not feasible due to exponentially growing compute
times and rounding errors \citep{deshpandeQuantumComputationalAdvantage2022a,bjorklund2019faster}. 

An essential task is to validate the output data of GBS in an efficient
and scalable way. Algorithms that replicate the GBS task \citep{bulmerBoundaryQuantumAdvantage2022a}
are not scalable to large sized networks, and checking large-scale
validity is necessary for quantum advantage (QA) claims \citep{drummondSimulatingComplexNetworks2022}.
Computers must be both fast and accurate. We show that matrix-P representations
projecting conserved parity can validate large-scale GBS experiments
in the low loss limit, where earlier phase-space methods are impractical.
Our approach complements studies of unitary ensembles \citep{drummond2016scaling,martinez2024linear,ehrenberg2025transition,ehrenberg2025second}
and anti-concentration \citep{ehrenberg2025second,ehrenberg2025transition},
by computing necessary, but not sufficient, tests of validity similar
to other RNG validation tests \citep{rukhin2010nist}.

The example treated is a calculation of binned count probabilities.
Other methods using loop Hafnian generator functions \citep{bulmer2024simulating}
are dominated by up to quartic scaling for the $10^{3}-10^{4}$ mode
networks of current interest. For these, matrix-P run times scale
nearly linearly with mode number, allowing much faster computation
of count probabilities. The matrix-P representation can also be used
for other tests \citep{drummond2016scaling,drummond2020initial,drummondSimulatingComplexNetworks2022}
in parallel. We investigate scalability up to $8000$ modes. This
size was reached in recent experiments \citep{liu2025robust}, although
not using PNR, and used in related QA proposals \citep{ehrenberg2025transition}.

We compare our results to the +P method, which can represent any nonclassical
state \citep{drummond1980generalised}. It has found many applications
\citep{carter1987squeezing,MunroReid:1993,Deuar:2007_BECCollisions,Kiesewetter2017pulsed,KiesewetterPhysRevA_2014,rosales2014probabilistic,Takata2015quantum,Drummond:2016,Teh_PRA2018},
including GBS experiments in current intermediate loss regimes with
threshold \citep{drummondSimulatingComplexNetworks2022,dellios2021,dellios2025validation}
and PNR \citep{dellios2025validationPNR} detectors. The end matter
shows +P converges too slowly for practical application to large,
low-loss PNR experiments.

\textbf{Matrix P-representations: }A phase-space representation \citep{Corney_PD_2003_GR_bosons,Corney_PD_PRL2004_GQMC_ferm_bos}
with a basis $\hat{\Lambda}(\vec{\boldsymbol{\alpha}})$ on can be
extended to a matrix representation by including a complete set of
$\mathcal{M}$ symmetry projectors, $\hat{\Pi}_{p}$, such that $\sum\hat{\Pi}_{p}=\hat{1}$
, where $p=0,\ldots\mathcal{M}-1$ indexes the global eigenvalues.
The quantum density matrix $\hat{\rho}$ is expanded in probabilities
$P_{\mathcal{M}}(\bm{\lambda})$ of coherent matrix projectors $\hat{\Lambda}_{pq}(\vec{\boldsymbol{\alpha}})=\hat{\Pi}_{p}\hat{\Lambda}(\vec{\boldsymbol{\alpha}})\hat{\Pi}_{q}$,
so that:
\begin{equation}
\hat{\rho}=\int P_{\mathcal{M}}(\bm{\lambda})tr_{\mathcal{M}}\left[\hat{\bm{\Lambda}}(\vec{\boldsymbol{\alpha}})\bm{\Omega}\right]\text{d}\bm{\lambda}.\label{eq:MCP_representation}
\end{equation}

Here $\bm{\lambda}=\left(\vec{\boldsymbol{\alpha}},\bm{\Omega}\right)$,
$\text{d}\bm{\lambda}\equiv\text{d}\vec{\boldsymbol{\alpha}}\text{d}\bm{\Omega}$,
while $\bm{\Omega}$ is an $\mathcal{M}\times\mathcal{M}$ effective
stochastic density matrix, and $tr_{\mathcal{M}}$ is the matrix trace.
The method treated here starts with the +P distribution, $P_{+}(\boldsymbol{\alpha},\boldsymbol{\beta})$,
with $\vec{\boldsymbol{\alpha}}=(\boldsymbol{\alpha},\boldsymbol{\beta})$,
which is positive and always exists. The matrix-P representation expands
in projected coherent states $\left\Vert \boldsymbol{\alpha}\right\rangle _{p}=\hat{\Pi}_{p}\left\Vert \boldsymbol{\alpha}\right\rangle $
\citep{Drummond2016coherent}, so that the matrix kernel $\hat{\bm{\Lambda}}$
is:
\begin{equation}
\hat{\Lambda}_{pq}(\vec{\boldsymbol{\alpha}})=\left\Vert \boldsymbol{\alpha}\right\rangle _{p}\left\langle \boldsymbol{\beta}^{*}\right\Vert _{q}e^{-w_{pq}\left(\vec{\boldsymbol{\alpha}}\right)},
\end{equation}
where $\left\Vert \boldsymbol{\alpha}\right\rangle =\exp\left(\boldsymbol{\alpha}\cdot\bm{a}^{\dagger}\right)\left|0\right\rangle $
is an unnormalized coherent state \citep{Glauber_1963_P-Rep}, with
$M$ bosonic creation operators $a_{k}^{\dagger}$ and an amplitude
vector $\bm{\alpha}$. We choose that $w_{pq}\left(\vec{\boldsymbol{\alpha}}\right)=\log\left(\sqrt{g_{p}(\vec{\boldsymbol{\alpha}})g_{q}(\vec{\boldsymbol{\alpha}})}\right)$,
where $g_{p}(\vec{\boldsymbol{\alpha}})\equiv\left\langle \boldsymbol{\beta}^{*}\right\Vert \hat{\Pi}_{p}\left\Vert \boldsymbol{\alpha}\right\rangle $,
so that $\hat{\bm{\Lambda}}$ has unit quantum trace and $\left\langle \bm{\Omega}\right\rangle $
is a density matrix.  One can also use other normalisations \citep{Carusotto:2001},
and the phase-variables $\boldsymbol{\alpha},\boldsymbol{\beta}$
can depend on the eigenvalues.

The expansion of Eq.(\ref{eq:MCP_representation}) unifies and extends
previous phase-space representations \citep{Glauber_1963_P-Rep,drummond1980generalised,Carusotto:2001,Deuar2006b}.
To prove that our expansion exists for all density matrices, we use
both completeness of the positive-P expansion and the projectors $\hat{\Pi}_{p}$.
Defining $n=\boldsymbol{\alpha} \cdot \boldsymbol{\beta}$, one can show that at least one positive matrix distribution 
exists such that
\begin{align}
P_{\mathcal{M}}(\vec{\boldsymbol{\alpha}}) & =P_{+}(\boldsymbol{\alpha},\boldsymbol{\beta})\prod_{pq}\delta\left(\Omega_{qp}-e^{w_{pq}-n}\right).\label{eq:canonical-expansion}
\end{align}
This is not unique, because coherent states are not orthogonal. The
advantage of this approach is to distinguish the $\mathcal{M}$ global
eigenvalues in each symmetry, which are finite and scalable, from
local, mode-dependent fluctuations in the exponentially large Hilbert
space basis for large mode number. Such local quantum effects are
described best by the coherent amplitudes in phase-space that can
then be randomly sampled efficiently. 

\textbf{Coherent state projections:} We utilize projectors that generate
Schr\"{o}dinger cat states \citep{yurke1986generating,Wolinsky_PRL1988,krippner1994transient,gravina2023critical,Drummond2016coherent}.
We define them to have $\mathcal{M}$ coherent amplitudes $\bm{U}^{q}\boldsymbol{\alpha}$,
where $q=0,\ldots\mathcal{M}-1$ and $\bm{U}$ is a unitary matrix
which is an $\mathcal{M}$-th root of unity, and defines the symmetry.
We treat phase symmetry with $\bm{U}^{q}=\exp\left(iq\phi\right)$
and $\phi=2\pi/\mathcal{M}$ in the results below. Other, more general
cases with different symmetries will be treated elsewhere.

The basis vector, $\underline{\left\Vert \boldsymbol{\alpha}\right\rangle }\equiv\left[\left\Vert \boldsymbol{\alpha}\right\rangle _{0},\ldots\left\Vert \boldsymbol{\alpha}\right\rangle _{\mathcal{M}-1}\right]^{T}$,
is:
\begin{align}
\left\Vert \boldsymbol{\alpha}\right\rangle _{p} & =\frac{1}{\mathcal{M}}\sum_{q=0}^{\mathcal{M}-1}e^{-ipq\phi}\left\Vert \bm{U}^{q}\boldsymbol{\alpha}\right\rangle .
\end{align}
 Each cat state has a different set of photon numbers. If the maximum
photon number is $N_{max}<\mathcal{M}$, the state $\left\Vert \boldsymbol{\alpha}\right\rangle _{p}$
has a fixed global photon number $p$. 

For dynamics, we define derivatives $\partial_{j}=\partial/\partial\alpha_{j}$,
together with global annihilation and creation matrices, $\bm{\sigma}$
and $\boldsymbol{\sigma}^{\dagger}$ where $\sigma_{p,q}=\delta_{p,q+1}^{(\mathcal{M})}$,
such that $\delta_{p,q}^{(\mathcal{M})}$ is a cyclic Kronecker delta
(modulo $\mathcal{M}$). This leads to the general operator identities
for any $\mathcal{M}$:
\begin{align}
a_{j}\underline{\left\Vert \boldsymbol{\alpha}\right\rangle } & =\alpha_{j}\bm{\sigma}\underline{\left\Vert \boldsymbol{\alpha}\right\rangle }\nonumber \\
a_{j}^{\dagger}\underline{\left\Vert \boldsymbol{\alpha}\right\rangle } & =\partial_{j}\bm{\sigma}^{\dagger}\underline{\left\Vert \boldsymbol{\alpha}\right\rangle }\label{eq:ket-identities}
\end{align}

In the simplest case of $\mathcal{M}=2$, both matrices are a Pauli
matrix, $\boldsymbol{\sigma}=\boldsymbol{\sigma}^{\dagger}=\sigma^{x}$.
Operator products give similar identities, for example, $a_{i}^{\dagger}a_{j}\underline{\left\Vert \boldsymbol{\alpha}\right\rangle }=\alpha_{j}\partial_{i}\underline{\left\Vert \boldsymbol{\alpha}\right\rangle }$.

Differential identities for the kernel $\hat{\bm{\Lambda}}$ require
the introduction of a diagonal matrix $\bm{T}$, where:
\begin{equation}
T_{pp'}=\delta_{pp'}T_{p}=\delta_{pp'}g_{p-1}/g_{p}.
\end{equation}
The indices are cyclic, so $g_{-1}=g_{\mathcal{M}-1}$. A super-matrix
$\mathcal{T}\hat{\bm{\Lambda}}\equiv\left(\bm{T}\hat{\bm{\Lambda}}+\hat{\bm{\Lambda}}\bm{T}\right)/2$
is also needed for the identities. Applying Eq.(\ref{eq:ket-identities}),
one obtains  
\begin{align}
a_{i}^{\dagger}a_{j}\hat{\bm{\Lambda}} & =\alpha_{j}\left[\partial_{i}+\beta_{i}\mathcal{T}\right]\hat{\bm{\Lambda}}\nonumber \\
\hat{\bm{\Lambda}}a_{i}^{\dagger}a_{j} & =\beta_{i}\left[\tilde{\partial}_{j}+\alpha_{j}\mathcal{T}\right]\hat{\bm{\Lambda}},\label{eq:Lambda-identities}
\end{align}
where we have defined $\tilde{\partial}_{j}\equiv\partial/\partial\beta_{j}$
. The above identities can also be used to calculate expectation
values of arbitrary observables $\hat{O}$ corresponding to c-number
matrix functions $\bm{O}\left(\vec{\boldsymbol{\alpha}}\right)$,
such that:
\begin{equation}
\left\langle \hat{O}\right\rangle =\int P_{\mathcal{M}}(\vec{\boldsymbol{\alpha}})tr_{\mathcal{M}}\left[\bm{O}\left(\vec{\boldsymbol{\alpha}}\right)\bm{\Omega}\right]d\vec{\boldsymbol{\alpha}}.
\end{equation}
In the simplest parity representation with $\mathcal{M}=2$, $T_{0}=\tanh\left(n\right)$,
$T_{1}=\coth\left(n\right)$.

\begin{figure}
\begin{centering}
\includegraphics[width=0.4\textwidth]{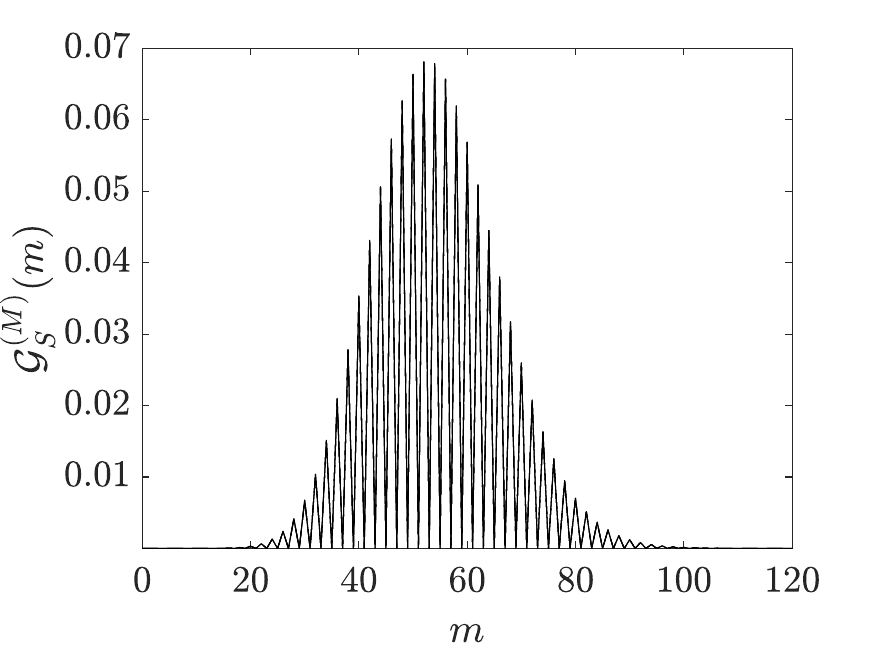}
\par\end{centering}
\caption{Lossless matrix-P simulations of $\mathcal{G}$, the total photo-count
probability in GBS, versus counts $m$. These are indistinguishable
from the exact distribution for uniform pure squeezed states with
$M=200$, $r=0.5$, and a Haar random unitary network. Errors are
 $<6\times10^{-5}$, and are not visible. An even larger size
of $M=10^{4}$ gives an error of $\sim6\times10^{-6}$, showing excellent
scalability. \label{fig:Exact_+C_PNR_comp}}
\end{figure}

\begin{figure}
\begin{centering}
\includegraphics[width=0.4\textwidth]{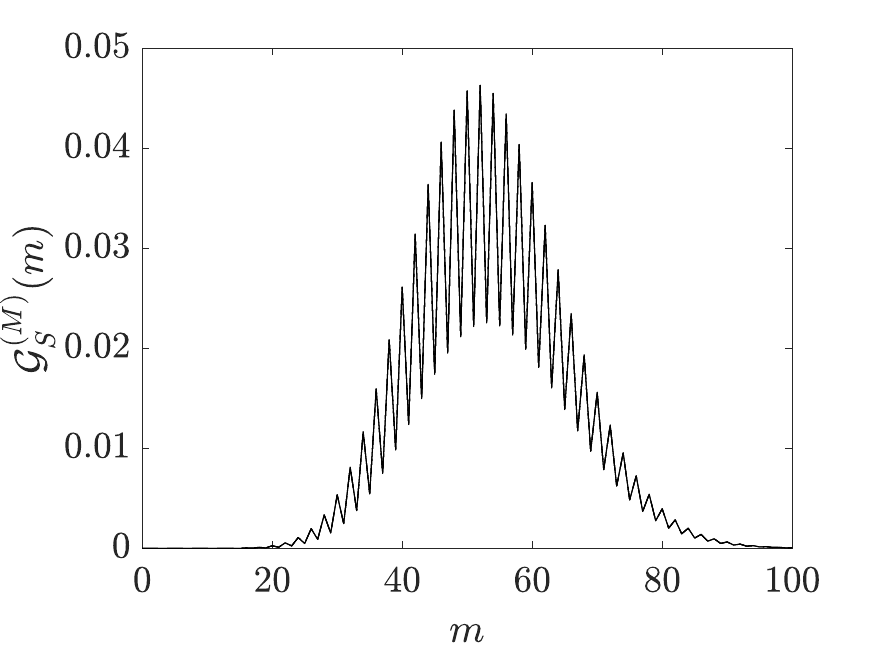}
\par\end{centering}
\caption{Simulations with 1\% loss, otherwise as in Fig (\ref{fig:Exact_+C_PNR_comp}).
\label{fig:Exact_+C_PNR_comp-lossy}}
\end{figure}

\textbf{Applications to GBS:} To demonstrate the advantages of matrix-P
in a topical case, we analyze the validation of lossless and low-loss
GBS with PNR detectors, using the approach of multiple statistical
tests \citep{rukhin2010nist}. In these regimes it is classically
hard to generate the output samples for large mode number \citep{Aaronson:2014,bulmerBoundaryQuantumAdvantage2022a,deshpandeQuantumComputationalAdvantage2022a}.
We show that in both lossless and lossy cases, very large scale validation
is feasible due to greatly reduced sampling errors. We treat one and
two-dimensional count binning, but other, low-order tests should be
used as well \citep{dellios2025validationPNR}. The resulting distributions
in Fig.(\ref{fig:Exact_+C_PNR_comp}), give excellent agreement with
exact results in soluble cases. 

Since squeezed states are always generated in correlated photon pairs,
the GBS examples treat a parity symmetry with $\mathcal{M}=2$, using
two-component cat states, $\left\Vert \boldsymbol{\alpha}\right\rangle _{p}=\frac{1}{2}\left[\left\Vert \boldsymbol{\alpha}\right\rangle +\left(-1\right)^{p}\left\Vert -\boldsymbol{\alpha}\right\rangle \right],$
so $\left\Vert \bm{U}^{q}\boldsymbol{\alpha}\right\rangle =\left\Vert \left(-1\right)^{q}\boldsymbol{\alpha}\right\rangle =\left\Vert \boldsymbol{\alpha}\right\rangle _{0}+\left(-1\right)^{q}\left\Vert \boldsymbol{\alpha}\right\rangle _{1}$.
Opposite parity states are orthogonal, because even parity ($p=0$)
states have only even photon numbers and odd parity ($p=1$) states
have odd photon numbers. 

To calculate GBS photonic dynamics, we treat a linear network with a Hamiltonian
$\hat{H}=\hbar\omega_{ij}a_{i}^{\dagger}a_{j}$. The equations are
identical to those in the positive P-representation, with:
\begin{equation}
\partial_{t}P_{\mathcal{M}}=i\omega_{ij}\left[\partial_{i}\alpha_{j}-\tilde{\partial}_{i}\beta_{j}\right]P_{\mathcal{M}}.
\end{equation}
This equation is solved using characteristics, so that:
\begin{align}
\left[\bm{\alpha}\left(t\right),\bm{\beta}\left(t\right)\right] & =\left[e^{-i\bm{\omega}t}\bm{\alpha}\left(0\right),e^{i\bm{\omega}t}\bm{\beta}\left(0\right)\right].
\end{align}
For a linear network, amplitude vectors are multiplied by a Haar random
unitary matrix $\boldsymbol{U}_{H}$, just as in the +P solutions
\citep{dellios2021}. Each coherent component evolves independently
of the others, as first pointed out by Schrödinger \citep{Schrodinger_CS}.
Damping is treated in detail below.

Next, take the GBS case, with squeezed vacuum input states $\left|\bm{r}\right\rangle $.
These are an integral over coherent states \citep{adam1994complete},
using an even function of $\bm{\alpha}$ defined as: $\xi\left(\bm{\alpha}\right)=\prod_{j}C_{j}e^{-\alpha_{j}^{2}\coth\left(r_{j}\right)/2}$,
with $C_{j}=\left[2\pi\sinh\left(r_{j}\right)\right]^{-1/2}$, for
real $r_{j}>0$. A squeezed state, $\left|\bm{r}\right\rangle $,
is an integral over a real line, $\left|\bm{r}\right\rangle =\int\xi\left(\bm{\alpha}\right)\left\Vert \boldsymbol{\alpha}\right\rangle d^{M}\bm{\alpha}$.
The +P distribution for pure squeezed states is defined on a $2M$
dimensional real subspace \citep{drummondSimulatingComplexNetworks2022}.
For multi-mode squeezed states $\left|\boldsymbol{r}\right\rangle \left\langle \boldsymbol{r}\right|$
with an $\mathcal{M}=2$ parity symmetric matrix-P method, one can
improve convergence enormously by noting that $\xi\left(\bm{\alpha}\right)$
is even, so that one can replace $\left\Vert \boldsymbol{\alpha}\right\rangle $
by the cat state $\left\Vert \boldsymbol{\alpha}\right\rangle _{0}$:
\begin{equation}
\hat{\rho}=\int\xi\left(\bm{\alpha}\right)\xi\left(\bm{\beta}\right)\left\Vert \boldsymbol{\alpha}\right\rangle _{0}\left\langle \boldsymbol{\beta}\right\Vert _{0}\delta\left(Im\left(\vec{\boldsymbol{\alpha}}\right)\right)d\vec{\boldsymbol{\alpha}}.
\end{equation}

The resulting matrix-P distribution, $P_{2}(\vec{\boldsymbol{\alpha}})$,
has $\Omega_{00}=1$, and is even in $n$. The +P sampling procedure
\citep{dellios2021} for initial stochastic trajectories can still
be used. The resulting output photon number sample is obtained from
matrix transformations $\boldsymbol{t}$ of the input amplitudes where
$n'=\sum_{i}n'_{i}=\sum_{i}\alpha'_{i}\beta'_{i}$ such that $\boldsymbol{\alpha}'=\boldsymbol{t}\boldsymbol{\alpha}$,
$\boldsymbol{\beta}'=\boldsymbol{t}^{*}\boldsymbol{\beta}$ are the
transformed amplitudes of a linear network, as one would expect classically. 

\textbf{Photon-counting observables:} Using Eq (\ref{eq:Lambda-identities}),
and taking the density matrix trace with the number operator, $\hat{n}_{j}=a_{j}^{\dagger}a_{j}$,
gives the result, on defining $p_{m}=\left[1-(-1)^{m}\right]/2=\left[0,1\right]$,
that photon-counting moments are given by:
\begin{align}
\left\langle :\hat{n}_{j}^{m}:\right\rangle  & =\int\left(\alpha_{j}\beta_{j}\right)^{m}tr_{\mathcal{M}}\left[\bm{T}^{p_{m}}\bm{\Omega}\right]P_{\mathcal{M}}(\vec{\lambda})\text{d}\vec{\lambda}.\label{eq:moment}
\end{align}
Apart from the weights, the even powers give expressions identical
to the +P distribution. The observables of interest for PNR detectors
are the $d$-th dimensional probability of binned number patterns
\citep{dellios2025validationPNR}
\begin{equation}
\mathcal{G}_{\boldsymbol{S}}^{(\tilde{n})}(\boldsymbol{m})=\left\langle \prod_{j=1}^{d}\left[\sum_{\sum c_{i}=m_{j}}\hat{G}(\boldsymbol{c})\right]\right\rangle ,\label{eq:General_GCP}
\end{equation}
where the projection operator for an $M$-mode output photon number
pattern $\bm{c}=[c_{1},\dots,c_{M}]$, where $c_{i}=0,1,2,\dots$
are observed photon counts, in the set $S$ is
\begin{equation}
\hat{G}(\boldsymbol{c})=\bigotimes_{i\in S}\frac{1}{c_{i}!}:\hat{n}_{i}^{c_{i}}e^{-\hat{n}_{i}}:.\label{eq:projection_op}
\end{equation}
Here, $m_{j}=\sum_{i\in S_{j}}c_{i}$ is the $j=1,\dots,d$-th photon
count bin, $\tilde{n}=\sum_{j=1}^{d}M_{j}\leq M$ is the correlation
order, and $\boldsymbol{S}=(S_{1},\dots S_{d})$ is a vector of subsets
of modes. 

We first treat the total count probability, which arises when $d=1$,
$\tilde{n}=M$, and $S=\left\{ 1,\dots,M\right\} $, such that:
\begin{equation}
\mathcal{G}_{S}^{(M)}(m)=\left\langle \sum_{\sum c_{i}=m}\hat{G}(\boldsymbol{c})\right\rangle .\label{eq:total_counts}
\end{equation}
In the case of lossless GBS with uniform pure squeezed state inputs,
an exact $M$-mode total count probability distribution is known \citep{huangPhotoncountingStatisticsMultimode1989,zhuPhotocountDistributionsContinuouswave1990}:
\begin{equation}
\mathcal{G}_{S}^{(M)}(m)=\begin{cases}
\binom{\frac{M}{2}+\frac{m}{2}-1}{\frac{m}{2}}p_S^{M/2}(1-p_S)^{m/2} & \text{even}\:m\\
0 & \text{odd}\:m
\end{cases},\label{eq:Exact_multi_mode_squeezed_state}
\end{equation}
where $p_S=1/(1+\bar{n})$ is the success probability for detecting
a photon with $\bar{n}=\sinh^{2}(r)$ mean photons per mode. With
uniform photon loss, an exact $M$-mode probability distribution for
uniform input squeezing also exists \citep{deshpandeQuantumComputationalAdvantage2022a,dellios2025validationPNR},
which provides a useful analytic test case. This shows the validity
of the method, which can readily model more realistic, experimental
networks. 

The projection operator Eq.(\ref{eq:projection_op}) is rewritten
in terms of odd and even powers
\begin{equation}
\hat{G}\left(\bm{c}\right)=:\left[C\left(\hat{n}_{S}\right)-S\left(\hat{n}_{S}\right)\right]\bigotimes_{i\in S}\frac{\hat{n}^{c_{i}}}{c_{i}!}:,
\end{equation}
where $\hat{n}_{S}=\sum_{i\in S}\hat{n}_{i}$, and we have abbreviated
$C\equiv\cosh$, $S\equiv\sinh$. This result also holds if the counts
$\hat{n}{}_{i}$ come from binned groups of channels. Using the matrix-P
method we find that the c-number function corresponding to $\hat{\bm{G}}\left(\bm{c}\right)$
for parity-symmetric inputs is:
\begin{equation}
G\left(\bm{m}\right)=g_{0}\left(n_{S},m\right)\prod_{i\in S}\frac{n_{i}^{c_{i}}}{c_{i}!}e^{-n_{i}},\label{eq:c-number-variable}
\end{equation}
where, defining parity as $\pi_{m}\equiv1-2*rem(m,2)=\pm1$:
\begin{equation}
g_{0}\left(n_{S},m\right)=\left(\frac{1+\pi_{m}e^{2\left(n_{s}-n\right)}}{1+e^{-2n}}\right).
\end{equation}

For the case that all channels are counted in the set $S$, so that
$n=n_{S},$ the c-number pattern projector for odd counts becomes
$g_{0}\left(2m+1\right)=0$, as expected, while the c-number pattern
projector for even counts is: $g_{0}\left(2m\right)=n^{2m}/[C\left(n\right)\left(2m\right)!].$

\textbf{Computational results and losses:} Calculations using the
matrix-P distribution are given in Fig.(\ref{fig:Exact_+C_PNR_comp})
for a $200$-mode lossless GBS and compared to the exact distribution
Eq.(\ref{eq:Exact_multi_mode_squeezed_state}). Results with $E_{S}=1.2\times10^{6}$
samples agree with the exact distribution with errors of $<5\times10^{-5}$,
in agreement with central limit theorem sampling error estimates \citep{opanchuk2018simulating}.
Extensions to an $8,000$-mode network, as in current threshold detection
experiments \citep{liu2025robust}, give similar relative errors.
All simulations were carried out with a publicly available MATLAB
software package, xqsim \citep{Drummond2025}.

We now compare this approach with the positive-P method, which is
useful for lossy cases and click detection \citep{dellios2025validation,drummondSimulatingComplexNetworks2022,drummond2020initial}.
To simulate total counts with the +P-distribution, one uses Eq.(\ref{eq:c-number-variable}),
except with a unit prefactor, $g_{0}=1$. For a sample number of $E_{S}=1.2\times10^{6}$,
the +P results in the lossless PNR case have extremely large sampling
errors due to a tail-heavy distribution of trajectories, as shown
in the end matter.

Losses are modeled by adding reservoir channels that are not excited
or counted, so that the network matrix becomes a lossy transmission
matrix $\bm{t}$. We compare simulations for this lossy case with
exact results in Fig. (\ref{fig:Exact_+C_PNR_comp-lossy}) for the
previous $200$-mode network, with an intensity loss of $1\%$. The
even-odd count oscillations are retained, but with reduced amplitude.
These simulations used $10^{3}$ sub-ensembles of $10^{4}$ trajectories
each. Typical probability errors were $5\times10^{-6}$ , indicating
an excellent fit to the exact, lossy distribution. For these ensemble
sizes, +P simulations have large sampling errors whenever the even-odd oscillations
are present, becoming accurate if parity conservation is violated
with large losses \citep{dellios2025validationPNR}, or if there are
only small mode numbers. The matrix-P method also gives correct results
with large losses, but there is less computational benefit in this regime. 

Another alternative, the loop Hafnian generator method \citep{bulmer2024simulating}
agrees with our results for total counts, but has at least cubic or
quartic scaling. We show in Appendix B that matrix-P timing scales
nearly linearly with mode number up to $M=8000$, making it faster
by orders of magnitude for experiments with $M>$1000. 

\textbf{Higher-dimensional binning:} Matrix-P methods can also model
higher-dimensional binning ($d\geq2$), where the outputs are divided
into multiple groups, denoted by the vector $\boldsymbol{S}$ in Eq.(\ref{eq:General_GCP}),
giving many independent validation tests. When $\tilde{n}=M$, the
photo-count distribution becomes a joint probability distribution
over $d$ count bins $\mathcal{G}_{\boldsymbol{S}}^{(M)}(m_{1},\dots,m_{d})$.
Unlike the $d=1$ case, if the network matrix is a Haar random unitary
no elementary higher-dimensional binning result is known, but the
matrix-P method is still applicable.

\begin{figure}
\begin{centering}
\includegraphics[width=0.4\textwidth]{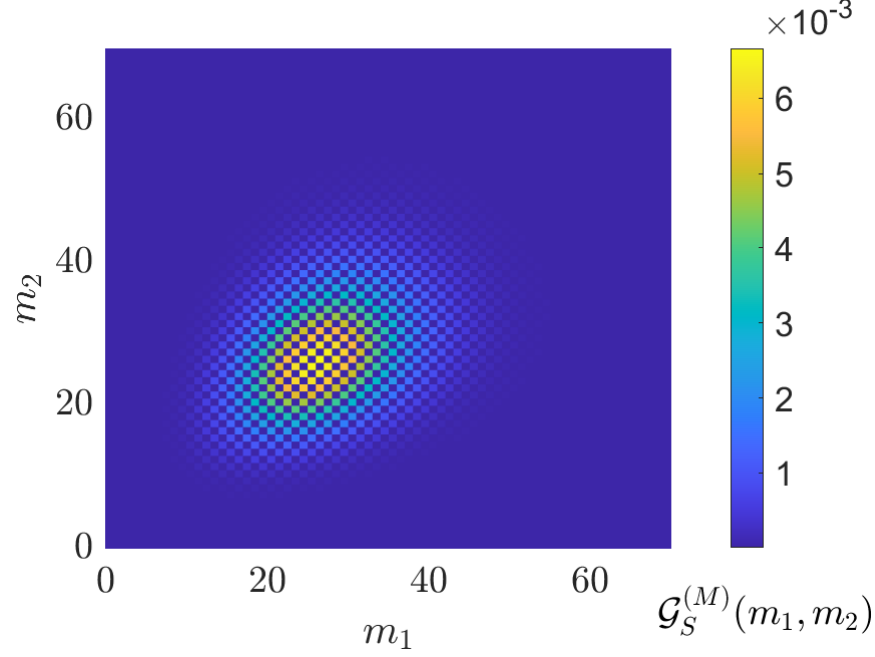}
\par\end{centering}
\caption{Surface plot of matrix-P simulated $d=2$ dimensional binned photo-count
probability $\mathcal{G}_{S_{1},S_{2}}^{(M)}(m_{1},m_{2})$, with
$S_{1}=\{1,\dots,100\}$ and $S_{2}=\{101,\dots,200\}$, for pure
squeezed states into a lossless GBS network. All parameters as in
Fig.(\ref{fig:Exact_+C_PNR_comp}). Sampling errors are $<3\times10^{-5}$.
\label{fig:Matrix_2D_binning}}
\end{figure}

In Fig.(\ref{fig:Matrix_2D_binning}), we simulate a $d=2$ dimensional
binning of a lossless $200$-mode network using the matrix-P distribution
with $E_{S}=1.2\times10^{6}$. The even-odd oscillations are still
present, with non-zero probabilities only occurring if $m_{1}+m_{2}$
is even. The sampling errors remain small at $<3\times10^{-5}$. Loop
Hafnian numerical results are not yet extended to higher-dimensional
binning for timing comparisons.

\textbf{Conclusion:} Rapid convergence is obtained for matrix-P simulations
of GBS photo-count distributions at large mode numbers. This will
allow precise validation of photon-counting data in future large-scale,
low-loss GBS quantum computing experiments, which are predicted to
have quantum advantage \citep{ehrenberg2025transition}. Improvements
occur because the method projects the physically relevant part of
the Hilbert space, allowing one to include any global symmetries.
This method allows other symmetries to be included, and has potential
applicability in many disparate areas, such as Bell violations \citep{rosales2014probabilistic},
ultra-cold atoms in optical lattices \citep{Carusotto:2001}, soliton
propagation \citep{Carter:1987,Drummond1993_Nature,ng2019nonlocal},
and others. 

\subsection*{Acknowledgements}

We thank Javier Mart\'inez-Cifuentes and co-authors for their Hafnian
code. This publication was made possible through an NTT Phi Laboratories
grant, and support of Grant 62843 from the John Templeton Foundation.
The opinions expressed in this publication are those of the author(s)
and do not necessarily reflect the views of the John Templeton Foundation.

\subsection*{Data Availability Statement}

All phase-space simulations were performed using the publicly available
software package, xqsim \citep{Drummond2025}, which is written in
the MATLAB programming language. 

\bibliographystyle{apsrev4-2}

\newpage{}

\clearpage{}

\section*{End Matter}

\paragraph*{Appendix A: Probability densities of observables -- }

Here we examine the distribution of stochastic observables for the
+P and matrix-P phase-space methods. 

The effect that conservation laws have on convergence of any observable
can be obtained from the probability density of the stochastic trajectories
for a specific observable. Trajectory amplitudes for any phase-space
observable $O(\boldsymbol{\alpha},\boldsymbol{\beta})$, with corresponding
operator $\hat{O}$, are binned into $N_{b}$ equally spaced bins
with spacing $\Delta=(O_{max}-O_{min})/N_{b}$ to estimate the probability
density as
\begin{equation}
P(O_{j})=\frac{1}{N_{S}\Delta_{b}}\sum_{i=1}^{N_{R}}\sum_{k=1}^{N_{S}}N_{j}^{(i,k)}.\label{eq:gen_probability_density}
\end{equation}
The sum is over the number of stochastic trajectories $N_{j}^{(i,k)}$
contained in the $j$-th bin. To improve sampling error estimates
\citep{drummondSimulatingComplexNetworks2022}, we use $N_{R}$ sub-ensembles
each with $N_{S}$ trajectories, giving an ensemble size of $E_{S}=N_{S}N_{R}$. 

\begin{figure}[h]
\begin{centering}
\includegraphics[width=0.8\columnwidth]{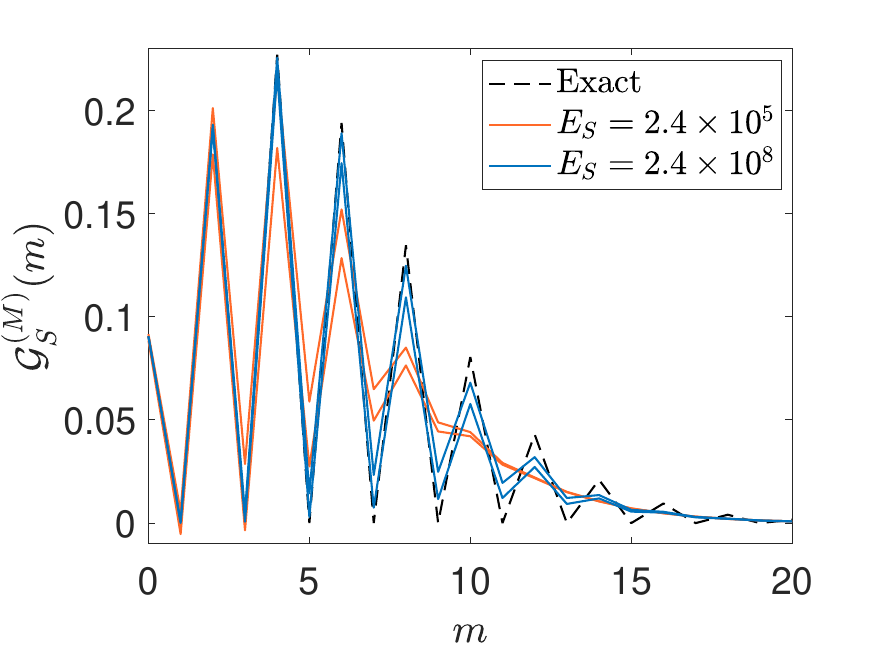}
\par\end{centering}
\caption{Comparisons of +P simulations of the total count distribution versus
the exact multi-mode photon counting distribution (dashed black line)
for lossless GBS with uniform $r=0.5$ and $M=20$. +P moments are
averages over $E_{S}=2.4\times10^{5}$ (solid orange line) and $2.4\times10^{8}$
(solid blue line) samples. Upper and lower lines correspond to $\pm1\sigma_{T,j}$
estimated sampling errors.\label{fig:Exact_+P_PNR_comp-1}}
\end{figure}

Our simulations in the main text used $N_{R}=120$ sub-ensembles of
$N_{S}=10^{4}$ trajectories each, giving maximum errors of $<6\times10^{-5}$
for the matrix-P photon-number probabilities, which are stochastic
observables. The +P method gave errors up to $\sim1000$ times larger
or more. To show the discrepancy, simulation results with $r=0.5$
and $M=20$ are given in Fig.(\ref{fig:Exact_+P_PNR_comp-1}). In
the lossless case with a Haar unitary, +P simulations have not yet
converged to the exact distribution, even with $E_{S}=2.4\times10^{8}$
samples.

To explain this, we computed the probability densities of stochastic
observables at different counts $m$ of the +P phase-space observable
$G_{+}\left(m\right)$ for a photon-number probability in a PNR type
GBS experiment, using 
\begin{equation}
G_{+}\left(m\right)=\frac{1}{m!}\left(n\right)^{m}e^{-n}.\label{eq:phase_space_projector}
\end{equation}

\begin{figure}
\begin{centering}
\includegraphics[width=0.5\columnwidth]{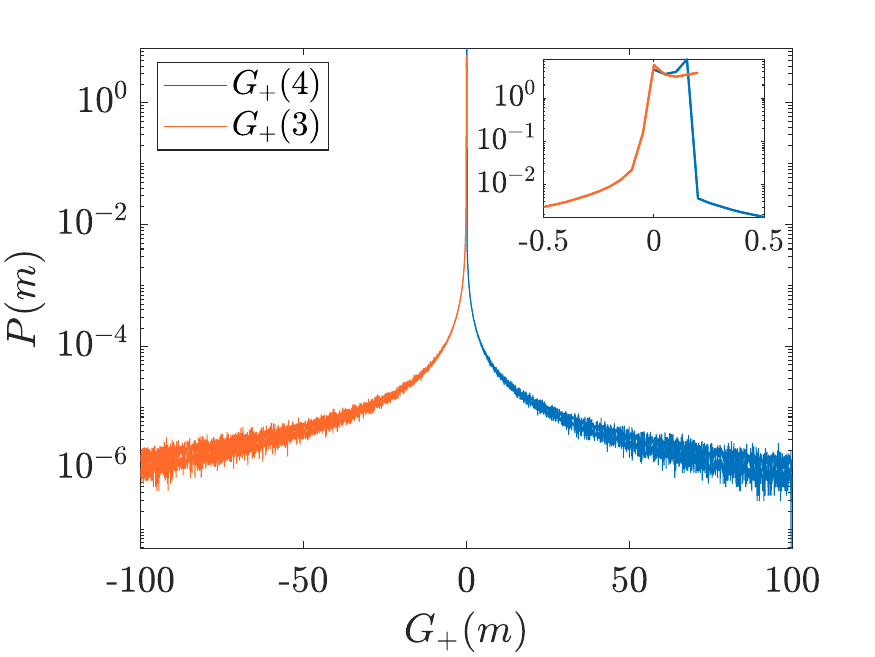}\includegraphics[width=0.5\columnwidth]{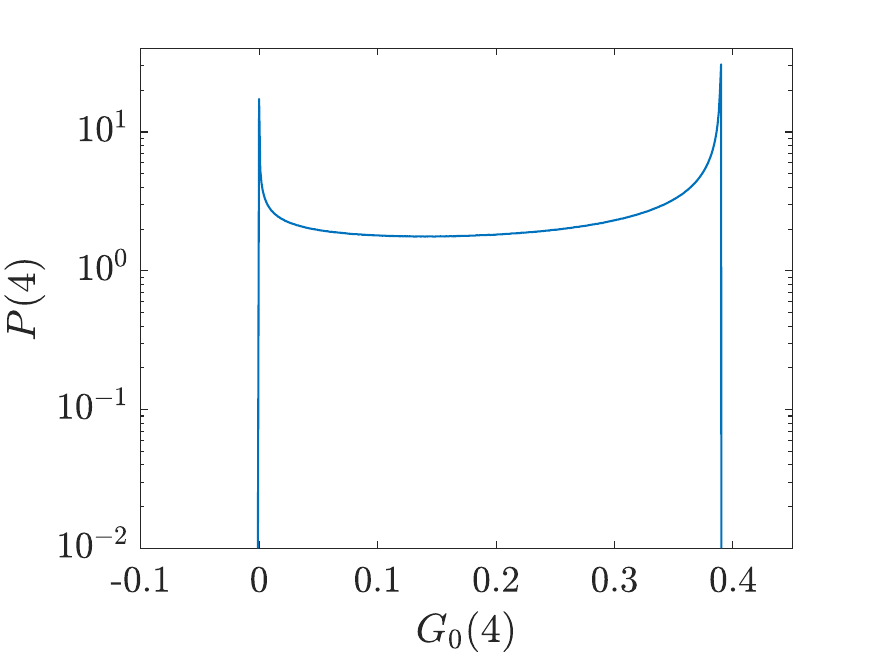}
\par\end{centering}
\caption{Left plot: Logarithmic plot of binned +P probability densities $P(m)$
versus observable $G_{+}(m)$ for $m=4$ (solid blue line), and $m=3$
(solid orange line). Trajectories are from simulations of a lossless
\textcolor{red}{${\normalcolor M=20}$} mode GBS network with $E_{S}=2.4\times10^{8}$,
using bins of size $\Delta_{b}=0.05$. The inset shows the range $[-0.5,0.5]$,
where the odd and even probability densities overlap. \protect \\
Right plot: Logarithmic plot of binned matrix-P probability densities
$P(4)$ versus observable $G_{0}(4)$ for the same trajectory and
mode count. Bins of size $\Delta_{b}=0.001$ are used to estimate
probabilities, with no probability tails. \label{fig:+P_m4_3_densities}}
\end{figure}

This demonstrates that the results in Fig.(\ref{fig:Exact_+P_PNR_comp-1})
occur because in the lossless +P case, the probability densities are
highly skewed with long tails, as shown in Fig.(\ref{fig:+P_m4_3_densities}),
where the $G_{+}\left(4\right)$ amplitudes have positive tails, while
$G_{+}\left(3\right)$ amplitudes have negative tails. The largest
probabilities in both cases are within the range $G_{+}\in[0,0.5]$
(see inset of Fig.(\ref{fig:+P_m4_3_densities})), giving an overlap
between the odd and even densities. 

The projector Eq.(\ref{eq:phase_space_projector}) used in the +P
simulations is unbounded due to the exponential in Eq.(\ref{eq:phase_space_projector}).
From the inset of the distribution in Fig.(\ref{fig:+P_m4_3_densities}),
the $G_{+}\left(4\right)$ values with the highest probability correspond
to observables with small amplitude. However, large positive trajectories
also occur, with small probabilities. To obtain correct results, these
tails must be sampled, requiring enormous numbers of samples. The
number of samples required for good accuracy grows rapidly as $M\rightarrow\infty$,
and even for the $M=200$ simulations in the main text, it is far
too large to be practical. 

By comparison, projecting onto the smaller Hilbert space of even photon
numbers makes matrix-P computation far more efficient. Due to the
parity representation expansion, odd counts are always zero in lossless
cases. In the right-hand panel of Fig.(\ref{fig:+P_m4_3_densities}),
we bin matrix-P amplitudes of the observable $G_{0}(4)$ for the total
photon count distribution of the $M=20$ mode network. Although accurate
convergence is obtained for smaller ensembles, we used $E_{S}=2.4\times10^{8}$
for comparison with the +P density estimates.

The matrix-P distribution does not have the large tails present in
the +P case. Instead, the distribution is highly localized and bimodal.
In these calculations, the matrix-P projector is within the range
$(0,0.4)$ for $m=4$, and is bounded for any mode number. This bimodal
matrix-P distribution has its largest probabilities at $G_{0}(4)\approx0.39$
and $G_{0}(4)\approx0$, with no extended tails.

In summary, once projected onto a smaller Hilbert space, the number
of samples required is reduced enormously, because the range of values
each trajectory can take is much smaller. This improvement is not
just for lossless cases, and the method is applicable to general states.
One must check that there are no other relevant conservation laws,
since these may also cause enhanced sampling errors if they lead to
heavy tailed distributions, but these too can be eliminated with projections. 

\paragraph*{Appendix B: Numerical scaling properties -- }

\begin{figure}[h]
\begin{centering}
\includegraphics[width=0.5\columnwidth]{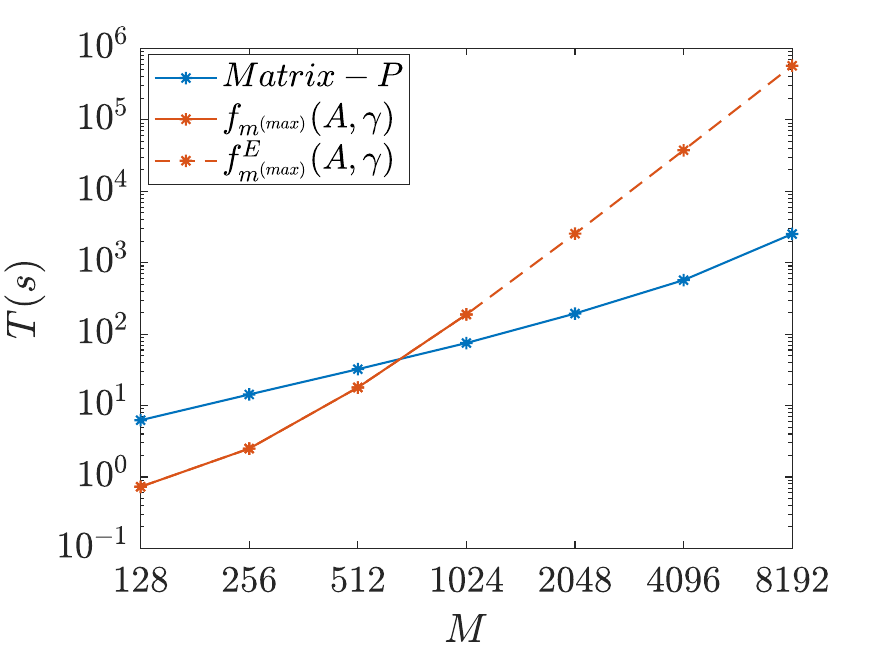}\includegraphics[width=0.5\columnwidth]{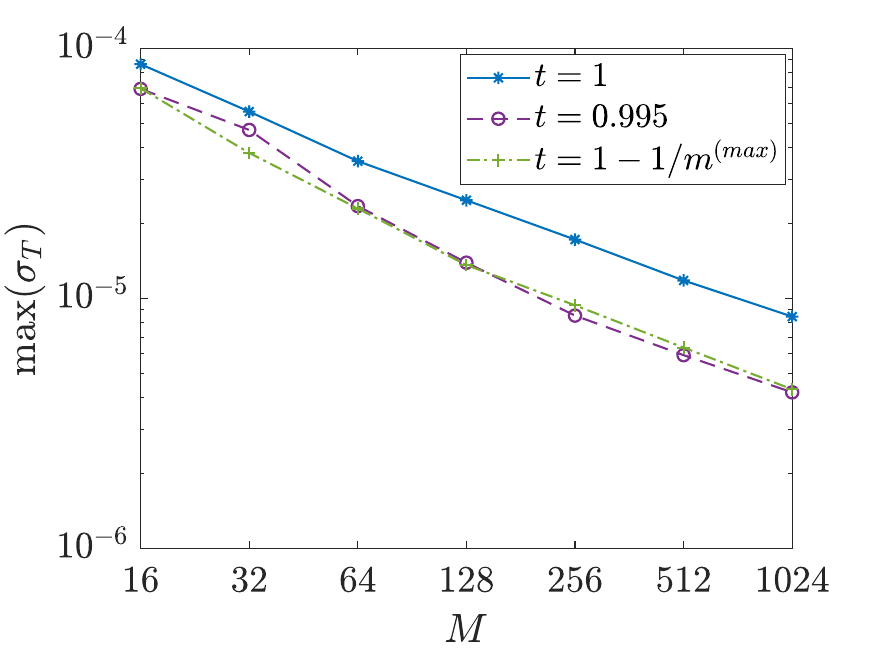}
\par\end{centering}
\caption{Scaling properties of matrix-P total count distribution simulations
using $E_{S}=1.2\times10^{6}$. Left plot: Run-times $T$ (seconds)
versus mode number $M$ of matrix-P (blue) and loop Hafnian $f_{m}$-function
algorithm (orange). Lossless GBS with uniform squeezing $r=0.89$
and a Haar unitary $\boldsymbol{U}_{H}$ is simulated for mode numbers
$M=128,256,..,8192$. The dashed orange line is the extrapolated $f_{m}$-function
run-time, denoted $f_{m^{(\text{max})}}^{E}$.\protect \\
Right plot: Scaling of the maximum matrix-P sampling error $\text{max}(\sigma_{T})$
versus $M=16,32,..,1024$ for uniform amplitude loss rates $t=1$
(solid blue), $t=0.995$ (dash purple), and $t=1-1/m^{(\text{max})}$
(dash dot green), applied as $t\boldsymbol{U}_{H}$, with nonuniform
squeezing $r_{i}=0.89+0.1w_{i}$ for the $i$-th input with Gaussian
random noise $w_{i}$. Here, $m^{(\text{max})}=85,118,174,273,450,776,1388$
for each network. \label{fig:Benchmarking}}
\end{figure}

Another method of computing binned count probabilities proposed by
Bulmer et al \citep{bulmer2024simulating} utilizes the loop Hafnian
generator function; the $f_{m}$-function. This useful development
gives results in agreement with ours. Here, we compare run times ($T$) versus
mode number, up to $M=8192$, of matrix-P and the $f_{m}$-function.
Scaling properties of the maximum matrix-P sampling errors, $\text{max}(\sigma_{T})$,
are also investigated when losses and nonuniform squeezing is present.
Results are shown in Fig.(\ref{fig:Benchmarking}).

Matrix phase-space methods scale generally as: $T\sim\mathcal{C}_{1}E_{S}M+\mathcal{C}_{2}E_{S}M^{2}+\mathcal{C}_{3}M^{2}$,
for an overall additive error $\sim1/\sqrt{E_{S}}.$ For the mode
numbers investigated here, the linear phase-space memory access term
$\mathcal{C}_{1}$ dominates (see Fig.(\ref{fig:Benchmarking})).
As $E_{S},M\rightarrow\infty$, the times should converge asymptotically
to the quadratic matrix multiply time contribution, $\mathcal{C}_{2}$.
Unlike the Hafnian approach, the same phase-space ensemble can be
used efficiently, to compute a large number of distinct moment or
marginal tests which are necessary for full validation.

The $f_{m}$-function has run-times estimated to scale as $T\sim c_{1}mM^{3}+c_{2}m^{2}\log\left(m\right),$
where $m$ denotes the photon count bin. At the time of writing, the
$f_{m}$-function algorithm has only been applied to total count distributions,
with one test at a time. Therefore we treat this case for comparative
timing, where the number of matrix-P samples is constant at $E_{S}=1.2\times10^{6}$. 

The largest observable count bin $m^{(\text{max})}$ is chosen to
occur when $\mathcal{G}_{S}^{(M)}(m^{(\text{max})})=10^{-7}$. This
cut-off is implemented in both the matrix-P and $f_{m}$-function
algorithms. With inputs $m^{(\text{max)}}\propto M$, as used here,
this gives $T\propto M^{4}$, while if $m^{(\text{max)}}\propto\sqrt{M}$
as in anti-concentration proposals \citep{ehrenberg2025transition},
one has $T\propto M^{3.5}$. To optimize the matrix-P algorithm, we
also define a minimum count bin $m^{(\text{min})}$, such that $\mathcal{G}_{S}^{(M)}(m^{(\text{min})})=10^{-7}$. 

The left-hand panel of Fig.(\ref{fig:Benchmarking}) graphs matrix-P
timing for lossless GBS networks of size $M=128,256,..,8192$ with
$r=0.89$. Both algorithms use $m^{(\text{max})}=273,450,776,1388,2562,4831,9267$,
while minimums of $m^{(\text{min})}=37,122,318,755,1678,3605,7561$
are used for matrix-P. Simulations were performed on a 12-core 3.7GHz
AMD CPU computer with 32GB of RAM. 

The $f_{m}$-function algorithm is faster than matrix-P for networks
in the range $16\leq M\leq512$ \citep{bulmer2024simulating}, however
it becomes slower once $M>512$ due to its quartic scaling in $M$.
Matrix-P times scale nearly linearly in $M$ for this range, making
it much more efficient for large networks. 

For networks with $M>1024$, the publicly available $f_{m}$-function
algorithm gave no useful results, due to floating point overflows
and underflows. For this reason, some of the run-times in Fig.(\ref{fig:Benchmarking})
were estimated (dashed orange line) by extrapolation. Using the $M\leq1024$
times, least squares minimization found the $c_{1}$, $c_{2}$ coefficients
in the $f_{m}$ scaling law, which were used to estimate times for
$M=2048,4096,8192$.

We now investigate the scaling of matrix-P sampling errors $\sigma_{T}\propto1/\sqrt{E_{S}}$
in experimentally realistic networks with losses and nonuniform squeezing.
Using $E_{S}=1.2\times10^{6}$ samples, networks of size $M=16,32,..,1024$
are simulated for randomly generated non-uniform squeezing parameters
$r_{i}=0.89+0.1w_{i}$ for the $i$-th input mode with Gaussian random
noise $w_{i}$. Results of $\text{max}(\sigma_{T})$ are presented
in the right-hand panel of Fig.(\ref{fig:Benchmarking}). Uniform
amplitude transmissions of $t=1,0.995$ are used, as well as $t=1-(m^{(\text{max})})^{-1}$,
where $m^{(\text{max})}$ is the same as the timing simulations for
each $M$. This loss scaling is chosen so that the even-odd oscillations
are still present in the total count distributions. Matrix-P sampling
errors  decrease for larger networks. 

In summary, quartic scaling causes the loop Hafnian $f_{m}$-function
run-times to increase rapidly. Matrix-P simulations of total count
probabilities for $M=8192$ are $\sim220$ times faster than the estimated
$f_{m}$-function run-time, and can treat multiple tests at once.
The ability to rapidly simulate large networks is vital to validation
of QC outputs in future experiments \citep{ehrenberg2025transition}.

\end{document}